# Imaging of objects through a thin scattering layer using a spectrally and spatially separated reference

Xiaoqing Xu,[1,2] Xiangsheng Xie,[1,3] Abhilash Thendiyamma,[2] Huichang Zhuang,[1] Junpeng Xie,[1] Yikun Liu,[1] Jianying Zhou,[1,4] And Allard P Mosk[2,*]

[1]*School of Physics, State Key Laboratory of Optoelectronic Materials and Technologies, Sun Yat-sen University, Guangzhou 510275, China*
[2]*Nanophotonics, Debye Institute for Nanomaterials Science, Utrecht University, P.O. Box 80.000, 3508 TA Utrecht, The Netherlands*
[3]*Department of Physics, College of Science, Shantou University, Shantou, Guangdong 515063, China*
[4]*stszjy@mail.sysu.edu.cn*
*[\*a.p.mosk@uu.nl](mailto:a.p.mosk@uu.nl)*

**Abstract:** Incoherently illuminated or luminescent objects give rise to a low-contrast speckle-like pattern when observed through a thin diffusive medium, as such a medium effectively convolves their shape with a speckle-like point spread function (PSF). This point spread function can be extracted in the presence of a reference object of known shape. Here it is shown that reference objects that are both spatially and spectrally separated from the object of interest can be used to obtain an approximation of the point spread function. The crucial observation, corroborated by analytical calculations, is that the spectrally shifted point spread function is strongly correlated to a spatially scaled one. With the approximate point spread function thus obtained, the speckle-like pattern is deconvolved to produce a clear and sharp image of the object on a speckle-like background of low intensity.

## References and links

1. I. Freund, "Looking through walls and around corners," Physica A **168**, 49–65 (1990).
2. A. P. Mosk, A. Lagendijk, G. Lerosey, and M. Fink, "Controlling waves in space and time for imaging and focusing in complex media," Nat. Photon. **6**, 283-292 (2012).
3. M. Gu, X. Gan, and X. Deng, *Microscopic imaging through turbid media* (Springer Berlin Heidelberg, 2015).
4. I. M. Vellekoop and A. P. Mosk, "Focusing coherent light through opaque strongly scattering media," Opt. Lett. **32**(16), 2309–2311 (2007).
5. O. Katz, E. Small, and Y. Silberberg, "Looking around corners and through thin turbid layers in real time with scattered incoherent light," Nat. Photon. 6, 549-553 (2012).
6. H. He, Y. Guan, and J. Zhou, "Image restoration through thin turbid layers by correlation with a known object," Opt. Express 21, 12539-12545 (2013).
7. C.-L. Hsieh, Y. Pu, R. Grange, G. Laporte, and D. Psaltis, "Imaging through turbid layers by scanning the phase conjugated second harmonic radiation from a nanoparticle," Opt. Express **18**, 20723-20731 (2010).
8. K. Si, R. Fiolka, and M. Cui, "Fluorescence imaging beyond the ballistic regime by ultrasound pulse guided digital phase conjugation," Nat. Photon. **6**, 657-661 (2012).
9. Y. M. Wang, B. Judkewitz, C. A. DiMarzio, and C. Yang, "Deep-tissue focal fluorescence imaging with digitally time-reversed ultrasound-encoded light," Nat. Commun. **3**, 928 (2012).
10. C. Ma, X. Xu, Y. Liu and L. Wang, "Time-reversed adapted-perturbation (TRAP) optical focusing onto dynamic objects inside scattering media," Nat. Photon. **8**, 931-936 (2014).
11. S. Popoff, G. Lerosey, M. Fink, A. C. Boccara, and S. Gigan, "Image transmission through an opaque material," Nat. Commun. **1**, 81 (2010).
12. S. M. Popoff, G. Lerosey, R. Carminati, M. Fink, A. C. Boccara, and S. Gigan, "Measuring the transmission matrix in optics: an approach to the study and control of light propagation in disordered media," Phys. Rev. Lett.**104**, 100601 (2010).
13. Y. Choi, C. Yoon, M. Kim, T. D. Yang, C. Fang-Yen, R. R. Dasari, K. J. Lee, and W. Choi, "Scanner-Free and Wide-Field Endoscopic Imaging by Using a Single Multimode Optical Fiber," Phys. Rev. Lett. **109**, 203901 (2012).
14. B. Judkewitz, R. Horstmeyer, I. M. Vellekoop, I. N. Papadopoulos and C. Yang, "Translation correlations in anisotropically scattering media," Nat. Phys. **11**, 684-689 (2015)

15. I. Freund, M. Rosenbluh, and S. Feng, "Memory Effects in Propagation of Optical Waves through Disordered Media," Phys. Rev. Lett **61**, 2328-2331 (1988).
16. S. Feng, C. Kane, P. A. Lee, and A. D. Stone, "Correlations and Fluctuations of Coherent Wave Transmission through Disordered Media," Phys. Rev. Lett.**61**, 834-837 (1988).
17. S. Schott, J. Bertolotti, J. F. Leger, L. Bourdieu, and S. Gigan, "Characterization of the angular memory effect of scattered light in biological tissues," Opt. Express **23**, 13505-13516 (2015).
18. J. Bertolotti, E. G. van Putten, C. Blum, A. Lagendijk, W. L. Vos, and A. P. Mosk, "Non-invasive imaging through opaque scattering layers," Nature **491**, 232-234 (2012).
19. O. Katz, P. Heidmann, M. Fink, and S. Gigan, "Non-invasive single-shot imaging through scattering layers and around corners via speckle correlations," Nature Photon. **8**, 784-790 (2014).
20. K. T. Takasaki, and J. W. Fleischer, "Phase-space measurement for depth-resolved memory-effect imaging," Opt. Express **22**, 31426-31433 (2014).
21. E. Edrei and G. Scarcelli, "Optical imaging through dynamic turbid media using the Fourier-domain shower-curtain effect," Optica **3**,71-74 (2016).
22. Y. Shi, Y. Liu, J. Wang, and T. Wu, "Non-invasive depth-resolved imaging through scattering layers via speckle correlations and parallax," Appl. Phys. Lett. **110**, (2017).
23. M. Cua, E. Zhou, and C. Yang, "Imaging moving targets through scattering media," Opt. Express **25**(12), 3935-3945 (2017).
24. T. Wu, J. Dong, X. Shao, and S. Gigan, "Imaging through a thin scattering layer and jointly retrieving the point-spread-function using phase-diversity," Opt. Express **25**, 27182-27194 (2017).
25. M. Hofer, C. Soeller, S. Brasselet, and J. Bertolotti, "Wide field fluorescence epi-microscopy behind a scattering medium enabled by speckle correlations," arXiv:1802.08986(2018).
26. J. A. Newman, and K. J. Webb, "Imaging optical fields through heavily scattering media," Phys. Rev. Lett **113**, 263903 (2014).
27. J. A. Newman, Q. Luo, and K. J. Webb, "Imaging Hidden Objects with Spatial Speckle Intensity Correlations over Object Position," Phys. Rev. Lett **116**, 073902 (2016).
28. H. Zhuang, H. He, X. Xie, J. Zhou, "High speed color imaging through scattering media with a large field of view," Sci. Rep. **6**, 32696 (2016).
29. E. Edrei and G. Scarcelli, "Memory-effect based deconvolution microscopy for super-resolution imaging through scattering media," Sci. Rep. **6**, 33558 (2016).
30. X. Xu, X. Xie, H. He, H. Zhuang, J. Zhou, A. Thendiyammal, and A. P. Mosk, "Imaging objects through scattering layers and around corners by retrieval of the scattered point spread function," Opt. Express, **25**(26), 32829-32840(2017).
31. S.Sahoo, D. Tang, and C. Dang, "Single-shot multispectral imaging with a monochromatic camera," Optica, **4**(10), 1209-1213 (2017)
32. Rebecca French, Sylvain Gigan, and Otto L. Muskens, "Speckle-based hyperspectral imaging combining multiple scattering and compressive sensing in nanowire mats," Opt. Lett. **42**, 1820-1823 (2017).
33. X. Xie, Y. Chen, K. Yang, and J. Zhou, "Harnessing the point-spread function for high-resolution far-field optical microscopy," Phys. Rev. Lett. 113(**26**), 263901 (2014).
34. X. Xie, H. Zhuang, H. He, X. Xu, H. Liang, Y. Liu and J. Zhou, "Extended depth-resolved imaging through a thin scattering medium with PSF manipulation" Sci. Rep. **8**, 4585 (2018).
35. N. Antipa, G. Kuo, R. Heckel, B. Mildenhall, E. Bostan, R. Ng, and L. Waller, "DiffuserCam: lensless single-exposure 3D imaging," Optica, **5**(1), 1-9(2018).
36. A. K. Singh, D. N. Naik, G. Pedrini, M. Takeda, and W. Osten, "Exploiting scattering media for exploring 3D objects," Light Sci. Appl. **6**, e16219, (2017).
37. S. Mukherjee, A. Vijayakumar, M. Kumar and J. Rosen, "3D imaging through scatterers with interferenceless optical system," Sci. Rep. **8**, 1134 (2018).
38. J. W. Goodman, *Speckle Phenomena in Optics: Theory and Applications* (Roberts & Co., 2007).
39. R. C. Gonzalez and R. E. Woods, *Digital Image Processing* (3rd Edition) (Prentice-Hall, Inc., 2006).
40. B. E. Bayer, U.S. Patent No. 3,971,065. Washington, DC: U.S. Patent and Trademark Office (1976).
41. J. R. Fienup, "Phase retrieval algorithms: a comparison," Appl. Opt. **21**, 2758–2769 (1982).

## 1. Introduction

Imaging through scattering media [1-3] is a challenge as the wavefront of the incident light is distorted by scattering due to the inhomogeneous distribution of the refractive index of the media. Several methods have been put forward to compensate for the wavefront distortion and facilitate imaging. Wavefront shaping [4-6] and optical phase conjugation [7-10] are widely used to compensate the heavy distortion caused by multiple scattering. Transmission matrix methods [11-13] are applied to obtain an image through scattering media. Speckle correlation [14] appears to be a very promising method relying on the so-called memory effect (ME), which is the angular tilt-invariant property of the speckle pattern when the incident light is rotated with a small angle [15-17]. Within the small ME range, the speckle pattern can be considered as the convolution of the point spread function (PSF) with the object's response. Methods based on speckle autocorrelation retrieval [18-25] can be exploited to reconstruct objects non-invasively. Coherent speckle-correlation [26, 27] methods provide field-based information through thick scattering media. Deconvolution methods [28-32] are also exploited to demonstrate fast, 3D, high-resolution and large field-of-view imaging. Among them, color imaging [28] is realized when R, G, B wavelength components of the object are retrieved from their corresponding PSFs captured by a color camera. Single-shot multi-spectral imaging has also been demonstrated [31] with a monochromatic camera. It was found that an image of an object can be reconstructed by deconvolution with the PSF obtained using different illumination, but only if the illumination spectra overlap. To realize multi-spectral imaging in Ref. [31] several spectral PSFs were recorded in advance, and PSFs at different wavelengths were assumed to be uncorrelated. Another method using a scattering medium and a microlens array with the measurement of the spectral intensity transmission is found to be capable of realizing hyperspectral imaging [32]. We have recently reported a deconvolution method to extend the depth of field (DOF) by scaling the reference PSF [33] according to the focal length of the imaging system [34]. Several other recent imaging techniques [35-37] have shown retrieval of depth information or even full 3D images using various methods to obtain the point spread function.

In this paper, we introduce a method to retrieve the PSF from the transmission pattern of a reference object that emits light at a different wavelength. The crucial ingredient is the presence of a strong correlation between the PSF with shifted wavelength, and the PSF with scaled spatial coordinates. The wavelength scaling can be combined with the spatial scaling methods of [34] to image incoherently illuminated or luminescent objects through a thin scattering layer, using a reference object that can be extended in size, have a non-overlapping spectrum and be in a different depth plane than the object being imaged. The PSF can be retrieved from a known, physically extended object, which is more robust and provides a higher SNR than the use of a point object in the case of incoherent illumination. We demonstrate the use of wavelength and depth scaling factors for imaging such separated objects. Furthermore, we make use of a color CCD camera to acquire two spectrally separated objects simultaneously and demonstrate single-shot reconstruction imaging. This method is practical and promising for real time imaging through dynamic scattering media. The principle we demonstrate here may be used to enable fluorescence imaging through turbid layers using reference objects emitting at a different wavelength.

## 2. Depth correlation and PSF scaling

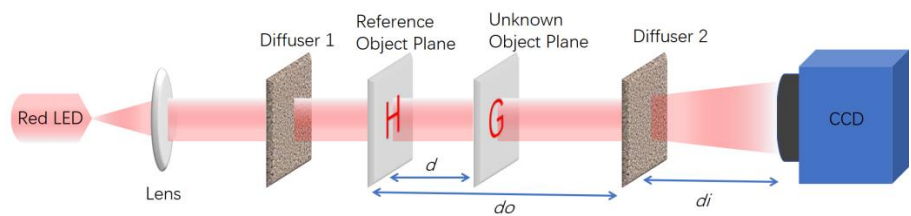

Fig. 1. Experimental setup: Imaging objects at another depth by deconvolution with modified PSF. The reference object and the unknown object are inserted alternately.

The experimental setup to demonstrate image reconstruction using depth correlation and PSF scaling is shown in Fig. 1. Light is emitted by a light emitting diode (LED) source (Daheng Optics,1W). The light is passed through two diffusers (diffuser 1, Newport 10-degree light shaping diffuser; diffuser 2, 5-degree) and the object to image is placed in between them. We make use of the a priori information of a reference object of a known shape for reconstructing the unknown object. The speckle pattern formed with spatially incoherent illumination has a lower contrast than that formed with a coherent source. However, such speckle patterns can be treated as the intensity sum of a series of coherent patterns from many coherence areas [38]. Here, different from our previous demonstration of depth-resolved imaging [34], we consider a more realistic situation in which the incident light is scattered by an additional layer. In this case, the exposure time of the CCD for obtaining the PSF from a point source reference has to be very large, which leads to a low signal to noise ratio (SNR) and is impractical in the case of dynamic scattering media. The use of an extended reference object generates a much brighter speckle pattern under the same illumination condition and permits the use of short exposure times. As shown in Fig. 2, a reference object with intensity distribution $O_R$ [a hollow letter 'H', Fig. 2(a)] is inserted between two diffusers and its speckle pattern $I_R$ [Fig. 2(c)] is recorded by the CCD (Basler ACA2040-90UM; pixel size, 5.5 μm). Within the memory effect range, the $I_R$ is the convolution of $O_R$ with the PSF, which can be expressed as $I_R = O_R * PSF$ (here, * represents convolution). Thus, the PSF can be retrieved by deconvolution from the reference speckle pattern $I_R$. Here, we chose the Wiener deconvolution algorithm [39], which yields

$$PSF = F^{-1}\left\{\frac{F\{O_R\}^*}{|F\{O_R\}|^2 + SNR^{-1}} F\{I_R\}\right\} . \tag{1}$$

Note that the speckle-like PSF can only be used to reconstruct unknown objects that are in the same plane with the reference object. If the unknown object [Fig. 2(b)] is inserted at a different depth plane, another speckle pattern $I_U$ is captured and within the memory effect range, $I_U = O_U * PSF'$ ($PSF'$ is the PSF for the unknown object's depth plane). Interestingly, if the scattering layer is thin, we can exploit a feature of the Fresnel diffraction regime, namely that the $PSF'$ can be approximately obtained by scaling the original PSF as follows:

$$PSF'(x, y) \approx m_z^2 PSF(m_z x, m_z y) , \tag{2}$$

$$\text{with} \quad m_z = \frac{(d_i + d_o)d_o'}{(d_i + d_o')d_o} , \tag{3}$$

where $m_z$ is the scaling factor [34]. Note that $m_z$ multiplies the coordinates of the PSF, so $PSF'$ is obtained by a coordinate stretch of $1/m_z$. The unknown object is reconstructed by a second deconvolution,

$$O_U = F^{-1}\left\{\frac{F\{PSF'\}^*}{|F\{PSF'\}|^2 + SNR^{-1}} F\{I_U\}\right\}. \tag{4}$$

In our experiment, the distance $d_o$ from the reference object to diffuser 2 is 117 mm and that from the unknown object plane to diffuser 2, $d_o' = 107$ mm. The CCD is at $d_i = 22$ mm from diffuser 2, leading to $1/m_z = 1.015$. In Fig. 2(a) and (b) we show the reference and unknown objects, respectively, and in Fig. 2(c) and (d) we show the respective speckle patterns. We retrieved the PSF [Fig. 2(e)] for the reference plane from the reference object's speckle pattern and rescaled it by $1/m_z$ to obtain the modified PSF [Fig. 2(f)]. A blurry image on a relatively high background [Fig. 2(g)] is reconstructed when the unknown object's speckle pattern is deconvolved with the retrieved reference PSF without any modification. On the other hand, a clear and sharp image on a much lower background [Fig. 2(h)] is achieved when the scaled PSF is used. To test the validity of the method, we moved the unknown object to another depth plane

$d_o$'=127 mm, which is 10 mm further from diffuser 2, leading to $1/m_z$=0.987. The corresponding speckle patterns and a clear reconstructed image are shown in Fig. 3. Thus, once a speckle pattern of reference object is obtained, other objects at different depth can be reconstructed by adjusting the scale of the reference PSF.

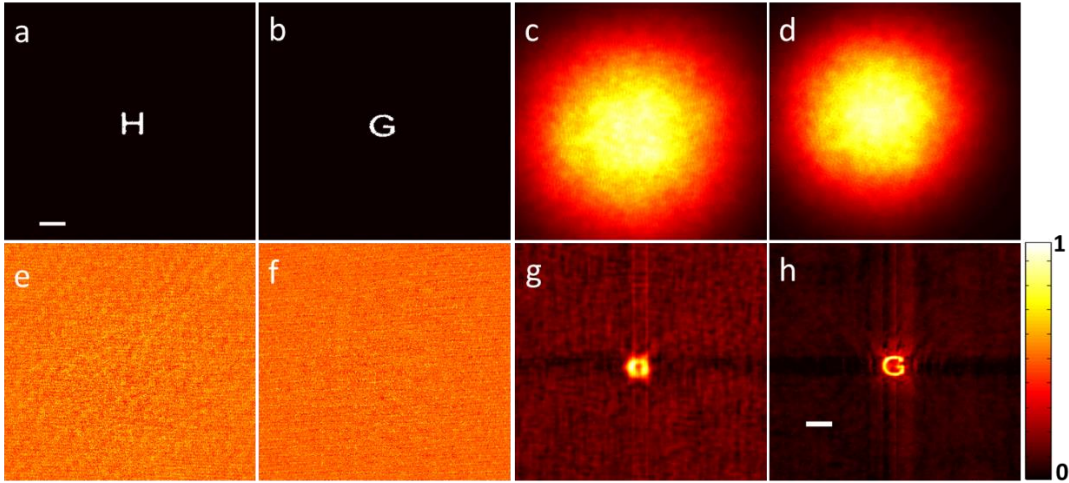

Fig. 2. Imaging objects at different depth with PSF scaling method and deconvolution. (a) is the reference object and the scalebar is 200 camera pixels and it is the same in Figs. (b-f). (b) is the unknown object. (c) and (d) are the speckle patterns of the reference object and unknown object respectively. (e) is the retrieved PSF from (c) and (f) is the modified PSF. (g) and (h) are the reconstruction using (e) and with (f), respectively. The scale bar in (h) is 50 camera pixels and it is the same in (g).

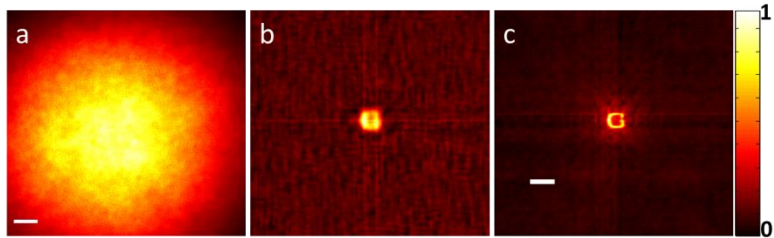

Fig. 3. Imaging objects at a different depth with PSF scaling method and deconvolution. (a) The speckle of unknown object at another depth. The scale bar is 200 pixels. (b) and (c) are the reconstruction from speckle pattern (a) with and without PSF scaling, respectively. The scale bar in (b) and (c) is 50 pixels.

## 3. Spectral correlation and PSF scaling

In most previous work based on deconvolution [28-30], the PSF and the speckle pattern of the unknown object are obtained under the same illumination source, which means that the spectrum of two speckle patterns is the same. Recently, it was reported that even though the PSF and the speckle pattern are obtained with two different broad band sources, deconvolution still works provided there is some spectral overlap, which was referred to as the cross-talk effect [31]. However, when using spectrally separated narrow-band light sources, objects cannot be reconstructed with the PSF corresponding to a different wavelength. Here we show that even in the spectrally separated case a reconstruction is possible. In this case the spatial coordinates of the reference PSF are scaled to correspond to a peak of the correlation function,

$$C(\lambda_1/\lambda_2, m) = \int dx\,dy\, PSF_{\lambda_1}(mx, my) PSF_{\lambda_2}(x, y) \quad . \tag{5}$$

An analytical calculation of this correlation function is presented in the Appendix. In Fig. 4(a) we show the shape of the correlation function for different wavelength ratios, for parameters that form a good approximation to our experimental conditions. It is seen that for the wavelength shifts of order a few percent that occur in the experiment there is a strong peak in the correlation function. In Fig. 4(b) we show the position of the correlation peak versus the wavelength ratio, for conditions that approach our experimental conditions, and for two different values of the correlation length of the diffuser. For a correlation length of 12.5 μm, which corresponds approximately to our 5-degree diffuser, we find the correlation peak occurs at $m_{peak} \approx (\lambda_1/\lambda_2)^{0.5}$. While convenient, the power law scaling is neither exact nor universal, the true algebraic scaling is given in the Appendix.

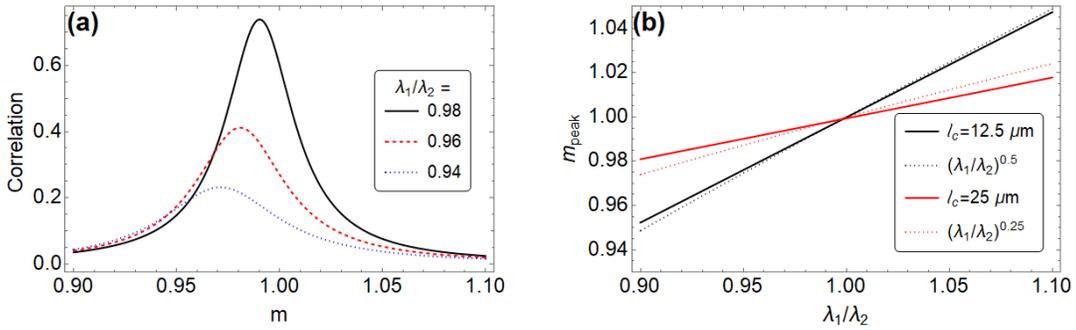

Fig. 4. (a) Normalized correlation between PSFs at different wavelengths as a function of the scaling factor m.
(b) Position of the maximum in the correlation function versus wavelength ratio, for two different correlation lengths of the diffuser.

Since $m$ scales the coordinates of $PSF_{\lambda 1}$, the approximation to $PSF_{\lambda 2}$ is obtained by a coordinate stretch of $PSF_{\lambda 1}$ by $1/m$. Therefore, an object with separated spectrum can be reconstructed with the corrected spectral PSF,

$$O_U = F^{-1}\left\{\frac{F\{PSF_{\lambda_2}\}^*}{|F\{PSF_{\lambda_2}\}|^2 + SNR^{-1}} F\{I_U\}\right\}. \tag{6}$$

The experimental setup to demonstrate imaging using spectral correlation and PSF scaling is shown in Fig. 5. Two objects (cutout letters "H" and "G") are illuminated by two LEDs (Daheng Optics, 1W) with different spectral profile [Fig. 6(a)]. The distance from the objects to diffuser $d_{o1} = d_{o2} = 180$ mm and that from the diffuser to the chip of the CCD ($d_i$) is 58.5 mm. Speckle patterns from these two objects are acquired by the CCD independently and the object is reconstructed by the deconvolution method with the PSF retrieved from the reference speckle pattern. In the second experiment, we introduced two narrow band filters before the LEDs to make sure the spectra of the light sources do not overlap [Fig. 6(e)]. The central wavelengths of the light sources with narrow band filters are 596nm and 620nm, respectively. Thus, the correlation maximum is expected for our experimental conditions at $1/m = 1.02$. The speckle patterns are obtained by inserting the corresponding narrow band filter before the CCD [Fig. 6(f) and (g)]. If the deconvolution method is used without scaling the PSF, only a blurry image is reconstructed [Fig. 6(h)]. However, upon scaling the retrieved PSF, a clear image is reconstructed [Fig. 6(i)]. This confirms the strong correlation between the spectrally shifted and the coordinate-scaled PSF enables us to obtain a clear image of the object.

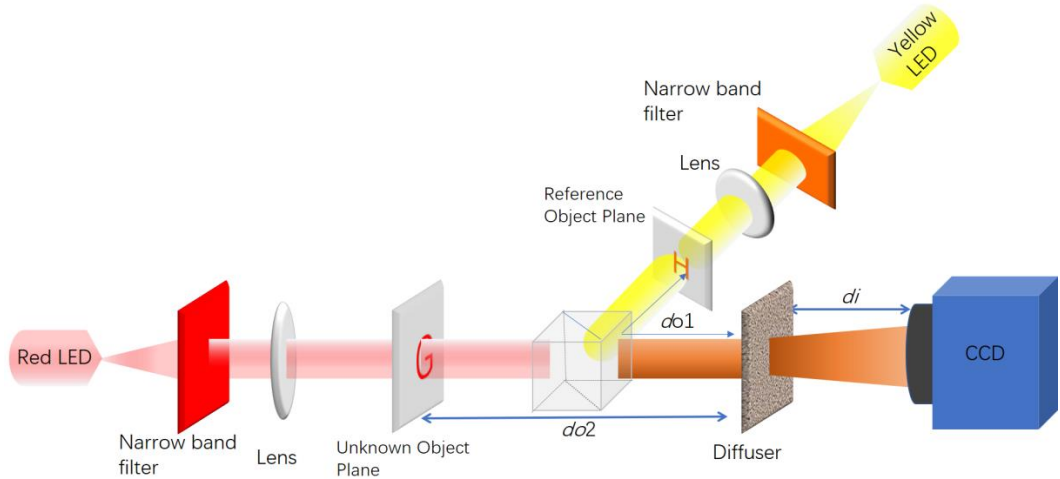

Fig. 5. Experimental setup to demonstrate reconstruction imaging using spectral correlation and PSF scaling.

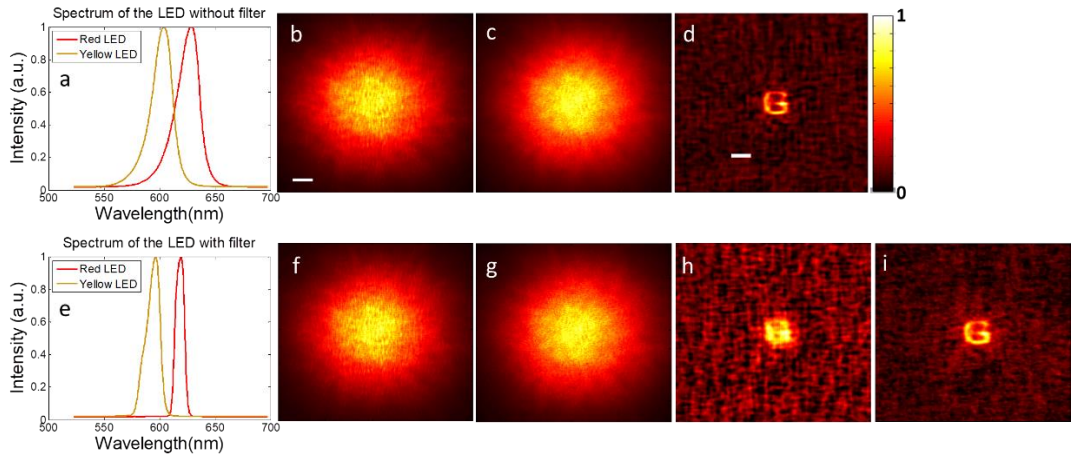

Fig. 6. Imaging objects with separated-spectrum at the same depth. The spectrum of LED source (a) without and (e) with filters. (b) and (c) are the speckle patterns of the reference object and unknown object under illumination without filters, and (d) is the reconstructed result. (f) and (g) are the speckle patterns of the reference object and unknown object under illumination with filter, and (h) and (i) are the reconstructed results without and with PSF scaling method. The scale bar in (b) is 200 pixels and it is same in (c), (f) and (g). The scale bar in (d) is 50 pixels and it is same in (h) and (i).

## 4. Spectral speckle pattern separation and single-shot imaging

In Sec. 2, the depth dependent PSF is obtained by scaling the PSF according to the reference depth whereas in Sec. 3, the wavelength dependence of the PSF scaling is depicted. When both depth and spectral differences are considered, the total scaling factor is given by

$$m_{total} = m_z \times m. \tag{7}$$

The unknown object "G" in Fig. 5 is moved 20 mm towards the diffuser and the two objects are now at different depths and illuminated by spectrum-separated sources with filters. In this case, $1/m_z$ is 1.03 according to Eq. 3. We use two corresponding narrow band filters to separate the speckle patterns of the reference object and unknown object, as in Sec. 3. The reference PSF is retrieved by deconvolution from the reference speckle pattern. The PSF corresponding to the object plane and wavelength is obtained by the appropriate coordinate stretch.

A very convenient method to obtain reference and object speckle patterns in a real time single-shot measurement is to use a color CCD. Thus the monochromatic CCD in Fig. 5 is replaced by a color one (acA2040-90uc - Basler ace). There is a Bayer filter [40] inside the camera in front of the photosensitive chip, which codes the wavelength response of each pixel. Using the spectral response curve of the CCD, different wavelength speckle patterns can be separated from one mixed speckle pattern. Suppose that there are two illuminating lights with different wavelength and their intensities are $I_{\lambda1}$ and $I_{\lambda2}$. For any pixel of the color CCD, R, G and B components of the captured speckle image [Fig. 7(a)] can be written as,

$$\begin{aligned} R &= R_1 \times I_{\lambda1} + R_2 \times I_{\lambda2} \\ G &= G_1 \times I_{\lambda1} + G_2 \times I_{\lambda2} \\ B &= B_1 \times I_{\lambda1} + B_2 \times I_{\lambda2} \end{aligned} \tag{8}$$

where $R_1$, $G_1$ and $B_1$ are the response coefficients of the CCD filter to the wavelength $\lambda_1$ and $R_2$, $G_2$ and $B_2$ are the counterparts to the wavelength $\lambda_2$. Note that there are three independent equations and only two unknowns in Eq. 8, so $I_{\lambda1}$ and $I_{\lambda2}$ can be found using pseudo-inverse methods. For our orange-red wavelength range simply using the first two equations suffices as the response of the B coded pixels is very low. Once $I_{\lambda1}$ and $I_{\lambda2}$ are obtained, a reference PSF is retrieved from $I_{\lambda1}$ (Eq. 1), and the PSF used to deconvolve the unknown object is achieved by rescaling the reference PSF with the total scaling factor $m_{total}$. The spectral sensitivity of the coded pixels is obtained from the sensor manufacturer. The center wavelengths of the filtered light sources are $\lambda_1 = 596$ nm and $\lambda_2 = 620$ nm and the corresponding (normalized) coefficients are $R_1$=0.67, $G_1$=0.29, $B_1$=0.04, $R_2$=0.88, $G_2$= 0.08, and $B_2$=0.04. The speckle patterns are calculated according to Eq. 8 and shown in Fig. 7(b) and (c). The reconstructed result is shown in Fig. 7(e). In order to test the scaling behavior, another object letter "F" [Fig. 7(d)] is used as unknown object and is reconstructed [Fig. 7(f)] from a different mixed speckle pattern.

We now turn to the effect of the spectral width of the sources on the recovery of the PSF. The sensor response for a spectrally broadened source is:

$$R = \overline{R_{\lambda 1}} \int I(\lambda_1) d\lambda_1 + \overline{R_{\lambda 2}} \int I(\lambda_2) d\lambda_2$$
$$G = \overline{G_{\lambda 1}} \int I(\lambda_1) d\lambda_1 + \overline{G_{\lambda 2}} \int I(\lambda_2) d\lambda_2 \quad , \tag{9}$$
$$B = \overline{B_{\lambda 1}} \int I(\lambda_1) d\lambda_1 + \overline{B_{\lambda 2}} \int I(\lambda_2) d\lambda_2$$

where $\overline{R_{\lambda 1}}, \overline{G_{\lambda 1}}, \overline{B_{\lambda 1}}, \overline{R_{\lambda 2}}, \overline{G_{\lambda 2}}, \overline{B_{\lambda 2}}$ are the average sensitivity coefficients of the CCD pixels weighted with the spectrum of the respective light sources. We measured the spectral response using the filtered LEDs. The R, G, B components of the acquired image can be obtained by adding all the corresponding pixel intensities, respectively. The ratio of the R, G, B components represents the sensor response to the light source and hence the spectrally averaged sensitivity coefficients. We find $\overline{R_{\lambda 1}}$ =0.6565, $\overline{G_{\lambda 1}}$ =0.3426, $\overline{B_{\lambda 1}}$ =0.0009, $\overline{R_{\lambda 2}}$ =0.8280, $\overline{G_{\lambda 2}}$ =0.0041, and $\overline{B_{\lambda 2}}$ =0.1679. The reconstructed results with the spectrally averaged sensitivity coefficients are shown in Fig. 7(g) and (h). These reconstructions have a higher signal to background compared to Fig. 7(e) and (f) which are reconstructed with coefficients for the central wavelength. Even the broad-bandwidth Bayer filters used in color imaging thus enable single-shot imaging of an object while simultaneously acquiring a spectrally separated reference.

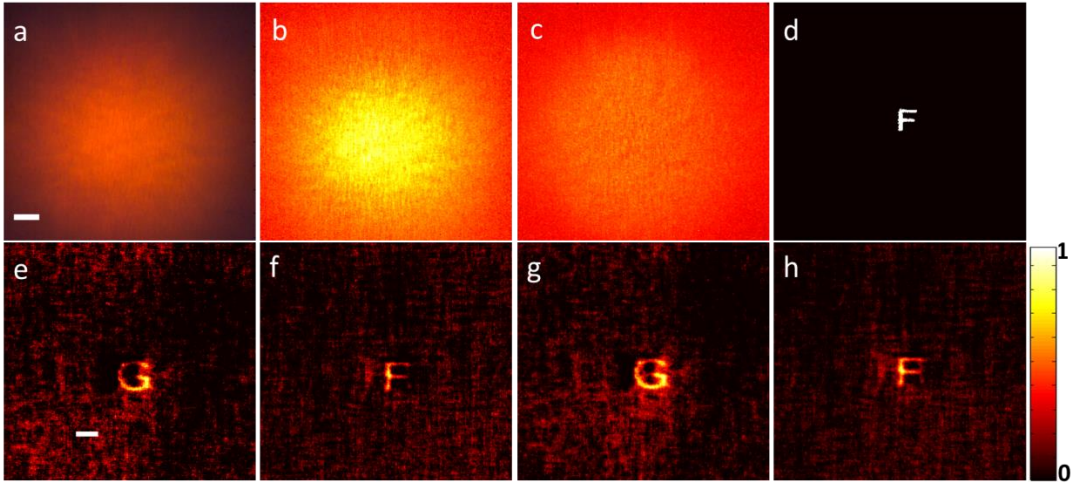

Fig. 7. Single-shot imaging of spectrum-separated objects at different depth. (a) is the hybrid speckle pattern of the reference object and unknown object. The scale bar is 200 pixels and it is the same in (a), (b), (c) and (d). (b) is the reference object's speckle pattern separated from (a) and (c) is that of the unknown object separated from (a). (d) is another unknown object. (e) and (f) are the reconstructed object, using the spectral sensitivities of the pixels at the center wavelength of the light sources. The scale bar is 50 pixels and it the same in (f), (g) and (h). (g) and (h) are reconstructed with the measured spectrally averaged sensitivity coefficients of the CCD filter.

## 5. Discussion

We have considered two scaling factors to resize the reference PSF for different depths and different wavelengths, and presented a single-shot speckle imaging method. The single-shot method is fast, robust and practical since the reference object can exist simultaneously with the unknown object. In the previous deconvolution-based works, the PSFs are formed by a point source and the unknown object is removed for capturing the PSF. Therefore, those techniques are not suitable to be used in dynamic scattering media as the PSF loses its correlation in time quickly.

We now discuss the field of view (FOV) and depth of field (DOF) attainable with our method. The FOV is limited by the optical memory effect, which is about $d_o \lambda / \pi L$ [15-17], where $L$ is the thickness of the scattering medium. For a holographic diffuser $L$ is effectively very small, on the order of a wavelength. The FOV can be enlarged by scanning methods [6] or by using an axial lens system to collect speckle patterns [28]. For large differences in depth and wavelength, the peak in the scaled correlation coefficient becomes much smaller than 1, as shown in Fig. 4(a), and as a result the deconvolution becomes imperfect. Unlike autocorrelation-based speckle scanning methods [18], our method is invasive requiring the presence of a reference object. However, it is possible to combine these techniques to realize non-invasive depth and spectral imaging. One way we can think of is to use the speckle-autocorrelation method to reconstruct one of the unknown objects. The reconstructed object can then be used as a reference object and the objects of interest can be retrieved by deconvolution with the modified PSF. In this way the iterative phase retrieval algorithm [41], which requires a threshold SNR to converge [19], is needed only once and subsequent images can be recovered more easily.

## 6. Conclusion

We have shown that objects can be imaged through a thin diffusing layer using a reference PSF, which can be extracted form a known object that is located in a different plane and/or emits at a different wavelength, than the object of interest. The key finding is that for a thin diffuser the wavelength-shifted PSF correlates strongly with a coordinate-scaled PSF. Analytical calculation yields the shape of the correlation function and the peak position, which depend on the correlation length of the scattering medium and on the size of the illuminated spot. The use of a spectrally separated reference object allows for single-shot imaging where the reference and object are recorded in a single exposure. This method has been implemented in an especially convenient way using the Bayer filter of a color camera to separate the reference and object patterns. The presented single-shot reconstruction method is fast and robust, which opens up prospects of imaging objects through thin but highly dynamic scattering media.

**Appendix**

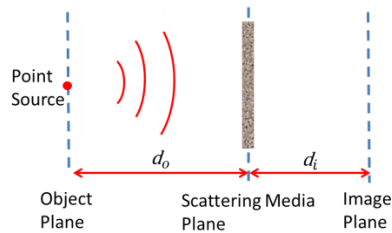

Fig. 8. Layout to calculate the PSF of a scattering imaging system

Here we present an analytical calculation of the correlation function between PSF's at different wavelength. Fig. 8 is the schematic of a scattering imaging system. We use a coordinate system where the diffuser is in the plane $z=0$. Assume we have a point source at location $(x,y,z)=-(d_0,0,0)$ with amplitude $S$. The field at the diffuser is $h(x,y,d_0)S$, where $h$ is the propagation function of vacuum, with $k=2\pi/\lambda$. The scattering effect by a thin diffuser can be modelled by a phase screen with transmission function $t(x,y)$. Thus the field just behind the diffuser is $t(x,y)h(x,y,d_0)S$. Neglecting a directional cosine, the field at the camera is found by integrating again with the propagation function,

$$E_i(x_i, y_i) = \int dxdy\ h(x_i - x, y_i - y, d_i)t(x, y)\ h(x, y, d_0)S\ . \tag{10}$$

We use the Fresnel approximation to the propagation function, introducing a Gaussian aperture of width $w$ at the diffuser,

$$h(x, y, d) = h_0 \exp(ik\frac{x^2 + y^2}{2d})\exp(-\frac{x^2+y^2}{w^2}). \tag{11}$$

To find

$$\begin{aligned}E_i(x_i, y_i) &= h_0 S \int dxdy\ t(x,y)\exp(ik\frac{x^2+y^2}{2d_0} + ik\frac{(x-x_i)^2+(y-y_i)^2}{2d_i} - \frac{x^2+y^2}{w}) \\ &= h_0^2 S\exp(ik\frac{x_i^2+y_i^2}{2d_i})\int dxdy\ t(x,y)\exp(ik\frac{x^2+y^2}{2f} + ik\frac{xx_i+yy_i}{2d_i} - \frac{x^2+y^2}{w})\end{aligned} \tag{12}$$

Here $f = d_o d_i /(d_o + d_i)$. The corresponding intensity, which is equal to the PSF, is

$$I_i(\rho_i, \lambda) = I_0 \int d^2\rho_1 d^2\rho_2 t(\rho_1)t^*(\rho_2) e^{-\frac{\rho_1^2+\rho_2^2}{w^2}} \exp\left(\frac{\pi i}{\lambda}\frac{\rho_1^2-\rho_2^2}{f} + \frac{2\pi i}{\lambda}\frac{(\rho_1-\rho_2)\rho_i}{d_i}\right). \tag{13}$$

Where $I_0$ is a prefactor and $\rho = (x, y)$. The correlation function can now be retrieved by filling the intensity in to Eq. (5) and performing the relevant integrations, noting that the integration over $\rho_i$ gives rise to a delta function, so that

$$\begin{aligned}C(n,m) &= C_0 \int d^2\rho_1 d^2\rho_2 d^2\rho_3 d^2\rho_4 t(\rho_1)t^*(\rho_2)t(\rho_3)t^*(\rho_4) e^{-\frac{\rho_1^2+\rho_2^2+\rho_3^2+\rho_4^2}{w^2}} \\ &\quad \exp\left(i\pi\frac{(\rho_1^2-\rho_2^2)+(\rho_3^2-\rho_4^2)/n}{\lambda f}\right)\delta\left((\rho_1-\rho_2)+(m/n)(\rho_3-\rho_4)\right)\end{aligned} \tag{14}$$

Where $n = \lambda_1 / \lambda_2$. We assume the diffuser is characterized by a correlation length $\ell_c$ so that ensemble averages of its transmission coefficient are given by

$$\langle t(\rho_a) t^*(\rho_b) \rangle = e^{-(\rho_a - \rho_b)^2 / \ell_c^2} \quad . \tag{15}$$

These correlations give rise to a direct term, which gives a smooth correlation background that we neglect, and a cross term that is strongly peaked. Normalizing the cross term so that $C(1,1) = 1$ we evaluate the integrals to find

$$\langle C(m,n) \rangle \approx \frac{2n^2 (f^2 \lambda^2 (\ell_c^2 + 2w^2) \ell_c^2 + \pi^2 \ell_c^4 w^4)}{f^2 \lambda^2 (\ell_c^2 + 2w^2)(\ell_c^2 (m^2 + n^2) + w^2 (m-n)^2) + \pi^2 \ell_c^2 w^4 (\ell_c^2 (m^2 + 1) + (m-1)^2 w^2)} \quad . \tag{16}$$

Examples are shown in Fig. 4a. The correlation function peaks at

$$m_{\text{peak}}(n) = \frac{f^2 \lambda^2 n w^2 (\ell_c^2 + 2w^2) + \pi^2 \ell_c^2 w^6}{(\ell_c^2 + w^2)(f^2 \lambda^2 (\ell_c^2 + 2w^2) + \pi^2 \ell_c^2 w^4)} \quad . \tag{17}$$

We fill in parameters that are comparable to the experimental situation, ($w$= 1 mm, $f$= 44 mm, $\lambda$= 600 nm, $\ell_c$=12.5 μm) to produce the results in Fig. 4(b).

## Funding

Netherlands Organization for Scientific Research (Vici 68047618); Chinese National Natural Science Foundation (11534017 & 61575223); China Scholarship Council (201606380037).

## Acknowledgments
The authors acknowledge S. Faez, J. Bosch and P. Pai for discussions.